\documentstyle[12pt]{article}
\makeatletter
\leftmargin - .1 in
\newcommand{\doublespacing}{\let\CS=\@currsize\renewcommand{\baselinesstrech}
{2.0}\tiny\CS}

\begin{document}
\setlength{\baselineskip}{18.5pt}
\centerline{\large{\bf Geometroneurodynamics}}
\vspace{0.9cm}
\centerline{\bf Sisir Roy$^{1,2}$, Menas Kafatos$^{2,3}$} 
\begin{center}
$^{1}${Physics and Applied Mathematics Unit, Indian Statistical Institute, \\ Calcutta-700 035, India,\\
e-mail: sroy@scs.gmu.edu}\\ 
\end{center}
\begin{center}
$^{2}${School of Computational Sciences, CEOSR, and \\
$^{3}$Department of Physics, George Mason University, \\
 Fairfax, VA 22030-4444, USA,\\ e-mail: mkafatos@gmu.edu}
\end{center}
\vspace{1cm}

 \abstract{\noindent{\small{Hilbert space structure is assumed as a valid geometric
description for neurodynamics, i.e., for applying any kind of quantum formalism 
in brain dynamics. The orientation selectivity of the neurons
is used as a justification to construct a type of statistical
distance function which is proportional to the usual distance
(or angle) between orientations of the neurons. The equivalence between 
the statistical distance and the Hilbert-space distance is discussed within 
this framework. It gives rise to the possibility of reanalysing the issue of 
measurement and information processing in the brain function.}}}
\vskip20pt
\noindent 

\newpage

\section{Introduction}
\noindent
Recent progress in brain research indicates (Stapp 1993, Penrose 
and Hemeroff 1996, Tegmark 2000, Pribram, 1991, Hagan et al. 2002) that to apply a wave function 
formalism used in quantum mechanics, it is necessary to consider Hilbert space structure.
 For that, one needs to consider first the geometric structure over the cortical 
surfaces of brain from the anatomical point of view (Roy and Kafatos 2003) and 
then its relation to Hilbert-space structure. To the best of our knowledge, no 
systematic attempt has been made so far to construct the geometric structure, 
starting from the neuronal characteristics over the cortical surface of brain and its 
connection to Hilbert space.  \\
\indent
Amari(2001) studied the geometrical structure of neuromanifold called 
multilayer perceptrons using information theoretic approach. In this approach, a family of neural 
networks constitutes of neuromanifold having probability distributions. Using Baysian approach, Amari constructed 
Reimannian metric tensor and applied this concept to the behavior of learning as 
well as statistical inferences over this kind of neuromanifold. It is worthmentioning 
that Pellionisz and Llin\'as(1982) and Llin\'as(2000) tried to understand the space-time 
representation for the internal world. According to their approach, cerebellum 
acts like a geometric tensor which connects the contravariant vector associated 
with the motor activities related to brain. A tensor network theory has been proposed by them so as to study 
the computational details associated to activities to brain. However, in this approach, 
it is quite difficult to define a smooth metric tensor globally i.e., over all the cortical surfaces of brain 
due to the existence of nonlinearities in some cortical surfaces. \\
\indent
The aim of this paper is to study the possibility of assigning any geometrical 
structure over the cortical areas of the brain. To do that, we first approach the problem considering  
neurodynamics from the physiological point of view and then attempts have been made 
to construct the corresponding Hilbert structure for the cortical surface of brain. 
The appropriate wave function for the neurons in the brain can be defined in 
Hilbert space as a geometric description of different cortical areas of the brain. 
The main fact lying behind the idea is that cells in different parts of brain, 
for example, the visual cortex exhibit orientation selectivity(Hubel 1995). 
The orientiation selectivity of the cells is  similar to  polarizing
filters producing a  beam of polarized photons in physics laboratory
experiments. \\
\indent
Also, we examine the notion of statistical distance as related
to  distinguishability of different oriented states of the cells 
in the cortical areas. It should be emhasized that we are not attempting 
to explain consciousness problem as such. Rather, we are trying to develop a possible theoretical 
framework within the quantum view to provide understanding for certain brain 
processes which do not appear to fit the classical paradigm. In section(2), we 
examine the orientation  selectivity of neurons as well as the relation between 
statistical distance and Hilbert space. In section(3), the spontaneous activity 
of neurons is examined. Finally, the possible implications of this work
are discussed in section (4).
 
\section{\bf Orientation Selectivity of Neurons and Statistical 
Distance }

There is a large variety as well as a large number of neurons in the brain.
 Collective effects which can only be accounted for in terms
of statistical considerations, are  clearly important. Up till now, experimental evidence
points to  more than 100 different types of neurons in the brain, although
the exact number is not yet fully known. It is found that no two neurons are
identical, and it becomes very difficult to say whether any particular
difference represents  more a difference  between  individuals or a difference
between different classes.
Neurons are often organized in clusters containing the same type of cell. The
brain contains thousands of cluster cell structures  which may take the
form of irregular clusters or of layered plates. One such example is the
cerebral cortex which forms a plate of cells with a thickness of a few
millimeters. In the visual cortex itself(Hubel and 1995), certain
clear, unambiguous patterns are found in the arrangement of cells with particular 
responses. \\
\noindent
 Even though our approach could apply to non-visual neurons, here
we limit our study to the neurons in the visual cortex as the visual cortex
is smoother preventing non-linear effects. For example, as the measurement 
electrode is moving at right angles to the surface through the grey matter, 
cells encountered one after the other have the same orientation as their 
receptive field axis. It is also found that if the electrode penetrates at an 
angle, the axis of the receptive field 
orientation would slowly change as the tip of the electrode is moved through 
the cortex. From a large series of experiments in cats and monkeys, it was 
found :  \\
\indent
{\it Neurons with similar receptive field axis orientation are located on 
top of each other in discrete columns, while we have a continuous change of 
the receptive field axis orientation as we move into adjacent columns}.

In the monkey striate cortex, about 70 to 80 percent of cells have the
property of orientation specificity. In a cat, all cortical cells seem to be
orientation selective, even those with direct genuculate input(Hubel,1995).
Hubel and Wiesel found a  striking difference among orientation-specific
cells, not just in the optimum stimulus orientation or in the position of 
the receptive field on the retina, but also in the way cells behave. The most useful 
distinction is between two classes of cells : simple and complex. The two types differ
in the complexity of their behavior and one can make the resonable assumption
that the cells with the simpler behavior are closer in the circuit to the input of the cortex.

The first oriented cell recorded by Hubel and Wiesel(Hubel,1995) which responded to the
 edge of the glass slide was a complex cell. The complex cells seem to have
 larger receptive fields than simple cells, although the size varies. Both type of cells
 do respond to the orientation specificity. There are certain other cells which 
respond not only to the orientation and to the direction of movement of the stimulus 
but also to the particular features such as length, width, angles etc. Hubel and Weisel
 originally characterized these as hypercomplex cells but it is not clear whether they 
constitute a separate class or, represent a spectrum of more or less complicated receptive fields. 

We now ask how the computational structure or filters can manifest as
orientation detectors ? Pribram(1981) discussed the question whether single
neurons serve as feature or channel detectors. In fact, Pribram and his
collaborators(1981,1991) made various attempts to classify "cells" in the
visual cortex. This proved to be impossible because each cortical cell 
responded to several features of the input, such as, orientation, velocity, and the
spatial and temporal frequency of the drifted gratings. Further, cells and cell
groups displayed different conjunctions of selectivities. From these findings and analysis, 
he concluded that cells are not detectors, that their receptive field properties could be
specified but the cells are multidimensional in their characteristics
(Pribram 1991). Thus, the pattern generated by an ensemble of neurons is 
required to encode any specific feature, as found  by Vernon Mountcastle's
work on the parietal cortex and Georgopoulos data(Pribram 1998) on 
the motor cortex. So the cells can merely be treated as filters.

Again, when dealing with the problem of perception, Freeman and his collaborators(1991) suggested
that perception cannot be understood solely by examining properties
of individual neurons i.e., by using the  microscopic approach that currently dominates
neuroscience research. They claimed that perception depends on the
simultaneous, cooperative activity of millions of neurons spread throughout
expanses of the cortex. Such global activity can be identified, measured and
explained only if one adopts a macroscopic view alongside the microscopic
building up view. 
\vskip5pt
\subsection{\bf Statistical Distance}
\noindent
To start with, we can define the notion of distance between the ``filters'' or the orientation
selective neurons in the similar manner i.e., to the statistical distance
between two quantum preparations, as introduced by Wootters (1981). The statistical
distance is most easily understood in terms of photons and polarizing filters : for 
example, let us consider a beam of photons prepared by a polarizing filter and analyzed
by a nicol prism. Let $\theta \in [0,\pi]$ be the angle by which the filter
has been rotated around the axis of the beam, starting from a standard
position ($\theta =0$) referring to the filter's preferred axis as being
 vertical. Each photon, when it encounters the nicol prism, has exactly two options : to
pass straight through the prism ( with  ``yes'' outcome) or to be deflected in
a specific direction characterised by the prism ( ``no'' outcome). If we assume
that the orientation of the nicol prism is fixed once and for all in such a
way that vertically polarized photons always pass straight through, then, 
by counting how many photons yield each of the two possible outcomes, an 
experimenter can learn something about the value of $\theta$ via the formula 
$ p= cos^{2} \theta$, where p is the probability of ``yes''(Wootters 1981), as
given by quantum theory.

If we follow this analogy in the case of oriented neurons in the brain i.e., as
if the filters are oriented in different directions like oriented analyzers,
we can proceed to define the statistical distance as shown in the 
following way: \\
Suppose the experimenter, in making his determination of the value of \ $\theta$,
has only a limited number of photons to work with, so that precisely n photons
actually pass through the filter to be analyzed by the nicol prism.
Then, because of the statistical fluctuations associated with a finite sample,
the observed frequency of occurence of ``yes'' is only an approximation to the
actual probability of ``yes'', the typical error being of the order of $\frac{1}{\sqrt n}$.
The experimenter's uncertainty in the value of $p$ is
\begin{equation}
 \delta p = {\left [\frac{p(1-p)}{n} \right ]}^{1/2}
\end{equation}
This uncertainty causes the experimenter to be uncertain as to the actual
 value of \ $\theta$ by an amount
\begin{equation} 
\delta \theta \ = \ {|\frac{dp}{d \theta}|}^{-1} \delta p \ =
\ {|\frac{dp}{d \theta}|}^{-1} {[ \frac{p(1-p)}{n}]}^{1/2}
\end{equation}
\vskip5pt
\noindent
In this way we can associate with each value of $\theta$, a region of
uncertainty  be  extending from \ $\theta - \delta \theta$ \ to \ $\theta + \delta 
\theta$.
Let us call two neighbouring orientations $\theta$ and $\theta\prime$
distinghuishable in $n$ trials if their regions of uncertainty do not overlap,
i.e, if
\begin{equation}
|\theta - \theta\prime | \geq {\delta \theta + \delta \theta \prime}
\end{equation}
If the uncertainty could be reduced to zero, one would effectively have an
infinite number of distinguishable orientations. This is , however, not the
case.

We now define the statistical distance $d(\theta_1,\theta_2)$ between
any two orientations $\theta_1$ and $\theta_2$ to be
\begin{equation}
d(\theta_1,\theta_2) = lim_{n \rightarrow \infty} \frac{1}{\sqrt n}[D]
\end{equation}  
where $D$ =  maximum number of intermediate orientations each of which
is distinghuishable (in $n$ trials) from its neighbours. Here, the statistical 
distance is obtained by counting the number of distinguishable states. This does 
not have anything to do, a priori, with the usual notion of distance
( or angle) between $\theta_1$ and $\theta_2$ which is, of course, $|\theta_1 - \theta_2|$.
However, it can be shown that these two types of distance are the same.
\noindent
From eqns(2)-(4) we can write the statistical distance in terms of $p(\theta)$
as follows:

\begin{equation}
d(\theta_1,\theta_2) \ = \ \frac{1}{\sqrt\pi}\int^{\theta_2}_{\theta_1} 
\frac{d\theta}{2 \delta\theta} \ = \ \int^{\theta_2}_{\theta_1}d\theta
\frac{|\frac{dp}{d\theta}|}{{2[p(1-p)]}^{\frac{1}{2}}}
\end{equation}

Substituing $p(\theta) = cos^{2}\theta$ \ we find
\begin{equation}
d(\theta_1 - \theta_2) = \theta_1 - \theta_2
\end{equation}
 which is equal to the angle between the two orientations. It should be noted that in 
order for the proportionality constant between the statistical distance and the actual 
distance to be unity, one needs to have

\begin{equation}
|\frac{dp}{d\theta}| \propto \left [ p(1-p) \right ]^{1/2}
\end{equation}
This will not be true if the probability law is different from
$$p(\theta) = cos^{2}\theta$$
Now, if we demand that the statistical distance be proportional to
$|\theta_1 - \theta_2|$, the $cos^2 \theta$  \ dependence of the probability
function must  necessarily follow and it is possible to define information measure 
according to above prescription. This can be reduced to Fisher information measure 
in the limiting case(Frieden, 1999). \\
\noindent
For example, if $W$ be Wootter's information 
measure and $I$ be the Fisher information measure, then 
\begin{equation}
W = cos^{-1} \left [\int d \theta p^{\frac{1}{2}} (\theta + \triangle \theta) \right ]
\end{equation}
\noindent
So,
$$W^2 \rightarrow \left [\frac{\triangle \theta^2}{4} \right ] I$$
Using the identities,
$$\int d \theta \ p^\prime(\theta) = 0 \ \ \ \ {\rm and} \ \ \ \ \int d \theta \ p^n(\theta) = 0$$
\noindent
where prime denotes the first and $n$ the $n$th derivative respectively. It is worthmentioning 
that Amari(2001) constructed the same distance function in neuromanifold using Fisher information 
measure but within Baysian framework where the metric concerned is considered to be of Reimannian 
nature. With these developments regarding the distance measure, the detail study of orientation 
selectivity of neurons might shed new light on the issue of information measure, suitable for the 
description of activity of brain and information processing. 

\subsection{\bf Statistical Distance and Hilbert Space}
\vskip5pt
\noindent
It can be shown(Wootters, 1981) that the statistical distance, as explained above, between 
two preparations, is equal to the angle in Hilbert space between the corresponding rays. The main
idea is as follows: \\
 Let us imagine the following experimental set up:  there are
two preparing devices, one of which prepares the system in a specific state, say
$\psi^1$, and the other prepares in $\psi^2$. Here, the statistical distance
between these two states can be thought as the measure of the number of distinguishable
preparations between $\psi^1 \& \psi^2$. However, in treating quantum systems,
new features should be observed as opposed to, like, just rolling the dice for a classical 
system. For dice, there is only one possible experiment to perform i.e., rolling the dice, where as
for quantum system there are many, i.e., one for each different analyzing device.
Furthermore, two preparations may be more easily distinguished with one anlyzing
device rather than with another. For example, the vertical and horizontal polarizations of 
photons can easily be distinguished with an appropriately oriented nicol prism, but can not
be distinguished at all with a device whose eigenstates are the right and left handed circular 
polarizations. Due to this reason, one can speak of the statistical distance between
two preparations $\psi^1 \& \psi^2$ to related to a particular measuring
device which means the statistical distance is device dependent. The absolute statstical
distance between $\psi^1 \& \psi^2$ is then defined as the largest
possible such distance i.e. statistical distance between $\psi^1 \& \psi^2$
when the states are anlyzed by the most appropriate or discriminating apparatus. \\
\indent
We can illustrate this point in the following way: \  Let $\phi^1, \phi^2 ..... \ \phi_N$ 
be the eigenstates of a measuring device $A$  by which $\psi^1 \& \psi^2$
are to be distinguished. It is assumed that these eigenstates are non-degenerate
so that there are $N$-distinct outcomes of each measurement. The probabilities
of various outcomes are \ $|(\phi^i,\psi^1|^2$ \ if the apparatus is $\psi^1 {\rm and} \ 
|(\phi^i,\psi^1|^2$, if the apparatus is $\psi^2$. Then the statistical distance
between $\psi^i \ \& \ \psi^(2)$ \ with respect to the analyzing device $A$ is
\begin{equation}
d_A{(\psi^1,\psi^2)} = cos^{-1} {\left [\Sigma_{i=1}|(\phi^i,\psi^1||(\phi^i,\psi^2| \right ]}
\end{equation}
This quantity attains its maximum value if one of the eigenstates of $A$
(say, $\phi^i$). In that case, we get the statistical distance as

\begin{equation}
d(\phi^1,\phi^2) = cos^{-1}\left |(\phi^1,\phi^2 \right |
\end{equation}
This clearly indicates that the statistical distance between two preparations is
equal to the angle in Hilbert space between the corresponding rays. This equivalence
between the statistical distance and the Hilbert space distance may appear very
surprising at first. But it gives rise to the interesting possibility that statistical
fluctuations in the outcome of mesurements might be partly responsible for Hilbert space
structure of quantum mechanics. In this way, these statistical fluctuations are as basic as the 
fact that quantum measurements are probabilistic in their nature.
\vskip5pt
\noindent
At this pint, a question may arise for the plausability of this approach for the following 
reason. It has been found that although representation of orientation of
objects in the visual cortex is fairly fine-scaled, visual information regarding
the nonstriate visual processing and in superior colliculus is very rough and
varies in a non-linear way from that in striate cortex. This has been supported by the 
recent findings from Van Essen Laboratory who have reconstructed cortical surface of 
brain from 3-$D$ data. However, for the present case, this kind 
of nonlinearity is neglected here, as we have adopted statistical consideration 
which averages out that kind of nonlinearity. And also, we consider the distance between the
different clusters of neurons or between the ensemble of neurons only. It is 
worthmentioning here that the statistical distance has been studied rigourosly 
also by Schweizer and Sklar(1983)in the context of cluster analysis, from  the mathematical 
point of view. 
\vskip5pt 
\noindent
\section{Concept of Measurement and Brain Functions}
\vskip5pt
\noindent
It is now clear that the statistical distance can be related to the angular distance in Hilbert 
space by applying the concept of measurement in quantum mechanics. We need to address the 
issue of measurement in the context of brain function so as to consider Hilbert space structure 
needed for any kind of quantum formalism. This is intimately related to the information processing 
in brain function. The information generated by integrated neural processes and its measurement 
remains one of the central issues of brain dynamics. The measure of information essentially depends 
on the basis of statistical foundation of information theory. One of the intriguing question arises 
is how far the statistical aspects of information theory can help one to assign a measure to 
differentiate the informative character of the neural processes without any reference to an external 
observer. The issue of external observer is debated in various branches of science and philosophy 
over the last century, since the birth of quantum mechanics. In the standard approach, one generally 
assigns a number to measure the information and probability of states of the system that are 
distinguishable from the point of view of an external observer. But the brain not only processes 
the information but also interprets the pattern of activities(Pribram 1991). Therefore, one must avoid 
the concept of privilaged viewpoint of an external observer to understand the information processing 
in the neural processes of the brain. \\
\indent
Edelman et al.(2000) discussed this problem in the context of 
neurobiology and consciousness. The main problem is how to measure the differences within a system , 
like the brain? He defined a concept of mutual information as considered in Shannon's framework. 
In this case, first a subensemable is considered and then the mutual information between this 
subensemable and and the rest of the states is found out. But, choosing the first subensemable, 
again, remains arbitrary. therefore, it is needed to analyze this situation in a more rigorous 
way so as to understand the measurement process in brain dynamics. \\
\indent 
Wright(1998), too, with the objectives to ascertain the minimal assumptions needed to reproduce the
experimental data, proposed a lumped continuum model of the cortex to explain EEG data
for synchronic oscillation in the cerebral cortex and object coherence. In his model, 
synchronicity depends wholly upon relatively long-range excitory connections in a continuum 
field-dendritic lag summations. Relatively rapid axonal transmission are considered as 
the essential ingredients while short range excitory/inhibitory interactions appear crucial 
only to the occurence of oscillation. These characteristics are very similar to that observed
in neural network models with interactional couplings observed by Wilson and
Bower(1991). The continuum formulation parallels and strengthens the neural network approach
by enabling a different insight into the physical nature of synchronicity,
since the essential nonlinearity of individual elements is avoided and the
stochastic and essentially linear properties of the neuronal mass is retained.
\vskip5pt
\noindent
On relatively long-range scales, the $cos^2{\theta}$ probability law follows from
the requirement that the statistical distance in case of neurodynamics is
proportional to the usual distance(angle) between orientations of a set of filters
or a set of neurons. It may be possible that in case of neurodynamics, the
statistical distance equals to the Hilbert space distance. Then we can think of 
Hilbert space structure over the cortical areas of the brain. Once this kind of Hilbert space structure
is envisaged over the cortical areas of the brain, it will be plausible to define quantum
processes to be valid underlying the neuronal dynamics.

\section{\bf Implications }
\vskip5pt
\noindent
It is evident from the above analysis that the law of probability related to
the statistical distance is similar to the {\it channel representation} as considered 
by Granlund(1999).

In the world around us, things generally appear different, whether they
irrespective of their exact form depending upon the different measuring parameters  with 
which it is looked through. Still we can recognize most objects at arbitrary positions, 
orientations, distances, etc. Some aspects of an object are  sufficiently familiar in order to begin 
the process of recognition even for arbitrary orientations. This may be the reason why one is interested
in the simultaneous appearance of similarity and difference in the properties
of objects. A representation of similarity requires a metric or characteristic distance
 measure between items to be defined. Granlund (1999) defined one type of such
distance which is related to channel representation. Here, each channel represents a particular 
property measured at a particular position of the input
space. Such a channel can be viewed in the analogy as the output from some band pass filter
sensor associated with a specific  property of the input. This relates the function of biological neural
feature channels. \\
\indent
In biological vision one can think of several examples
for the properties like edge, line or orientation detectors. If we consider
the channel output as derived from a  band pass filter, we can establish a measure
of distance or similarity in terms of the properties of the filter.
For  this channel representation, Granlund (1999) considered the measure for 
the output of the channel as $cos^2{\theta}$ where $\theta$ denotes 
the orientation of the filter. Of course he did not derive this analytically.
On the other hand our framework is consistent with his view. 
In conventional linear simple band pass filter, the phase distance between
the flanks is a constant times  $\pi/2$. Different filters will have different
bandwidths, but we can view this as a standard unit of similarity or
distance, related to a particular channel. Such a channel filter has the
characteristic that it is located in some input space, as well as local
in some property space. \\
\indent
The recent debates about the applicability of the wave function formalism in quantum mechanics 
as well as the relevance of quantum coherence for the information processing in the brain in 
various regions like visual cortex, auditory regions, olfactory bulbs etc. have raised lots 
of interest among scientific community. Pribram (1991) developed the first holographic model 
of the brain. According to
Pribram, the brain encodes information on a three dimensional energy field that
enfolds time and space, yet allows to recall or reconstruct specific images from
the countless millions of neurons, stored in a space slightly smaller than a melon. Together 
with his collaborators, he studied the implications of Gabor's
quanta of information for brain's function and their relation to Shannon's measure on the amount
of information contained in the data, obtained in their investigations. Rigorous investigations 
have been carried out(Jibu et al., 1996) further so as to study how the quantum mechanical processes 
can operate at the synaptodendritic level. \\
\indent
We have already discussed that Wootter's measure of information (and distance) is related to 
Fisher information measure which is considered as the mother of all information 
measure(Frieden, 1999) including that of Shannon's measure. So, their approach regarding 
information processing in brain should be reanalyzed using the concept of statistical 
distance function. However, arguments and counter arguments(Tegmark, 2000, Hameroff, 2002) have 
been raised recently about the applicability of the kind of model which, one way or other, 
is related to the application of quantum mechanical concepts.It is generally argued that 
the brain is warm and wet. Recent theoretical and experimental papers(Tegmark, 2000) support the 
prevailing opinion that the large warm systems will rapidly lose quantum coherence and 
classical properties will emerge out as a result. This rapid loss of coherence(decoherence problem) 
would naturally be expected to block any crucial role of quantum theory in explaining the 
interaction between our conscious experiences and the physical activities of our brain. \\
\indent
However, to mention further, in the quantum theory of mind  developed by Stapp(2000a), based 
on a relativistic version of von Neumann's quantum theory, the efficacy of mental effort is 
not affected by decoherenece problem. Briefly, Stapp(2000b) relates mind-brain to two
separate processes:  First, there is the unconscious mechanical brain process
governed by the Schrodinger equation which involves processing units that are
represented by complex patterns of neural activity (or more generally, of brain
activity) and, subunits within these units that allow "association" i.e., each unit
tends to be activated by the activation of several of its subunits.
The mechanical brain evolves by the dyanmical interplay of these associative units.
Each quasi-classical element of the ensemble that constitutes the brain creates, on the basis 
of clues, or cues, coming from various sources, a plan for having a possible coherent 
course of action. Quantum uncertainties entail that a host of different possibilities will emerge.
This mechanical phase of the processing already involves some selectivity, because
the various input clues contribute either more or less to the emergent brain
process according to the degree to which these inputs activate, via associations,
the patterns that survive and turn into the plan of action. \\
\indent
The issue of quantum coherence and consciousness has been discussed in detail, also by Hameroff and 
Penrose(1996) in their model, known as Orchestrated Objective Reduction(OrchOR) model, where the 
possible scenario for emergence of quantum coherence leading to OOR and conscious events is cellular 
vision. On the otherhand, Jibu et al.(1994) argued that this kind of processes require quantum 
coherence in microtubules and ordered water, based on the experimental observations(Albrecht-Buehler,
1992) which states that single cells utilize their cytokeletons in cellular vision-detection, 
orientation and directional response to beams of red/infrared light supports.

It is clear from the above models of Stapp, Penrose et al. that they used
the quantum mechanical wave function to describe both unconscious mechanical
brain processes connected with  the complex patterns
of neuronal activity as well as the conscious activity of the brain.
In both the approaches, a  Hilbert structure has been assumed in order to 
produce an evolution equation like Schrodinger equations. We like to emphasize that our approach 
 gives rise to a new possibility to construct the Hilbert space structure over the cortical surface 
of brain from an anatomical perspective. However, the issues, like concept of measurement, the role 
of observer and information measures should be thoroughly analyzed in the context of brain before 
applying any kind of quantum mechanical formalism.
\vskip20pt
\noindent 
{\bf Acknowledgements} \\
One of the authors (S.Roy) greatly acknowledges Center for Earth Observing and Space Research, 
School of computational Sciences, George Mason University, Fairfax, VA, USA, for their kind hospitality and 
funding for this work.
\vskip20pt
\noindent
{\bf References}

\begin{enumerate}

\item  Amari S.(2001), {\it IEEE Transaction on Information Theory},{\bf 47},1701-1711.
\item  Albrecht-Buehler G.(1992),{\it Rudimentary form of Cellular Vision},
       Cell. Biology, {\bf 89},8288-92.
\item  Donald M.J.(1990),{\it Quantum Theory and  the Brain}, Procd.Roy.Soct.
       London,{\bf A 427},43-53.
\item  Freeman W J.(1991),{\it The Physiology of Perception},Scientific American
       {\bf 264}, 78-85.
\item  Grandpierre A.(1995),{\it Quantum-Vacuum Interaction Dynamics in the 
       Brain} in The Interconnected Universe, by Ervin Laszlo, World Scientific, Appendix I,113-118.
\item  Grandpierre A.(1999),{\it The Nature of Man-Universe Connection},The Noetic Journal,{\bf 2},
       52-66.
\item  Granlund Goesta H.(1999),{\it  The Complexity of Vision},Signal 
       Processing,{\bf 74},101-116..
\item  Hagan S, Hameroff S. and Tuszynski J.A.(2002),Phys.Rev.{\bf E},{\bf 65},061901.
\item  Hameroff  S.and Penrose R.(1996),{\it Conscious Events as Orchestrated
       Space-time Selections}, Journal of Consciousness Studies ,{\bf 3},36-53.
\item  Hubel H.David(1995),{\it Eye, Brain, and Vision}(Scientific American
       Library, NY).
\item  Jibu M. , Hameroff S.,Pribram K H.,Yasue K.(1994),{\it Quantum Optical 
       Coherence in Cytoskeletol Microtubules : Implication
       for brain function},Biosystems {\bf 32},195-209. 
\item  Jibu M., Pribram K H.,Yasue K.,(1996),{\it From Conscious Experience
       to Memory Storage and Retrieval : The Role of Quantum Brain Dynamics and
       Boson Condensation of Evanescent Photons},Int.J.Mod.Phys.B,{\bf 10},1735-54.
\item  Lindahl B.I.B and Arhem P.(1994),{\it  Mind as a Force Field},J.Theort.Biology 
       {\bf 171},111-118 and references there in.
\item  Llin\'as R.(2002),{\it I of the Vortex}, MIT Press.
\item  Pellionisz A. and Llin\'as R.(1982), {\it Space-time Representation in the Brain Neuroscience},
       {\bf 7},2949-2970.
\item  Penrose Roger(1994),{\it Shadows of the Mind }(Oxford Press, Oxford, U.K.).
\item  Pribram K., Lassonde M.,Ptito M.(1981),{\it Classification of Receptive
       Field Properties in Cat Visual Cortex}, Exp.Brain Res.{\bf 43},119-130.
\item  Pribram K.(1991),{\it Brain and Perception- Holonomy and Structure in
       Figural Processing} ( Lawrence Erlbaum Associates Publishers, NJ).   
\item  Pribram K.(1998),{\it The History of Neuroscience in Autobiography},
       {\bf 2},335.
\item  Schweizer B. and Sklar A.(1983),{\it Probabilistic Metric Spaces}(North Holland)
\item  Roy Sisir(1998),'{\it Statistical Geometry and Applications to Microphysics
       and Cosmology}(Kluwer Academic Publishers).
\item  Spinelli D N., Pribram K H., Bridgeman B.(1970),{\it Visual Receptive Field
       Organization of Single Units in the Visual Cortex of Monkey},Int.J Neurosci,{\bf 1},67-70. 
\item  Stapp Henry(1993),{\it Mind, Matter and Quantum Mechanics}(Springer-Verlag,NY.Inc).
\item  Stapp  Henry(2000 a),{\it Decoherence, Quantum Zeno Effect and the 
       Efficacy of Mental Effort}, quant-ph/0003065
\item  Stapp Henry (2000 b),{\it From Quantum Nonlocality
       to Mind-Brain interaction }, quant-ph/0009062)
\item  Van Essen  Laboratory : cis.jhu.edu/wu\_research/brain.html
\item  Von Neumann John(1958),{\it The Computer and the Brain}(Yale University Press,1979).
\item  Wilson M A., and Bower J M.,(1991),{\it A computer Simulation of 
       Oscillatory Behaviour in  Primary Visual Cortex}, Neural Computation {\bf 3},498-509. 
\item  Wootters W.K.(1981),{\it Statistical Distance and Hilbert Space}, 
       Phys.Rev D {\bf 23},357-365.
\item  Wright J J.,(1998),{\it Simulation of EEG : Dynamic Changes in Synaptic
       Efficacy, Cerebral Rythms, and Dissipative and Generative Activity 
       in the Cortex},(Biological Cybernatics Press).
\end{enumerate}

\end{document}